# A First-principles Prediction of Two-Dimensional Superconductivity in Pristine B$_2$C Single layer

**Jun Dai, Zhenyu Li,\* Jinlong Yang,\* and Jianguo Hou**

**Hefei National Laboratory for Physcial Science at the Microscale, University of Science and Technology of China, Hefei, Anhui 230026, China**

*[a] Fax: 86-551-3603748; E-mail: zyli@ustc.edu.cn; jlyang@ustc.edu.cn.*

**Based on first-principles lattice dynamics and electron-phonon coupling calculations, B$_2$C sheet is predicted to be a two-dimensional (2D) phonon-mediated superconductor with a relatively high transition temperature (T$_c$). The electron-phonon coupling parameter calculated is 0.92, and it is mainly contributed by low frequency out-of-plane phonon modes and electronic states with a π character. When the Coulomb pseudopotential $\mu^*$ is set to 0.10, the estimated temperature T$_c$ is 19.2 K. To be best of our knowledge, B$_2$C is the first pristine 2D superconductor with a T$_c$ higher than the boiling point of liquid helium.**

Superconductivity in layered materials has attracted numerous research interests since the discovery of high transition temperature (T$_c$) cuprate. MgB$_2$[1] and iron-based superconductors[2] are two recent examples of layered superconductors. Although the mechanism of superconductivity in cuprates and iron-based superconductors are still under debate, MgB$_2$ is known to be a conventional superconductor, which can be well interpreted in the framework of Bardeen-Cooper-Schrieffer (BCS) theory.[3] Graphite intercalated compounds (GICs) is another type of layered BCS superconductor. Inspired by the relatively high T$_c$ found in YbC$_6$ (6.5 K) and CaC$_6$ (11.5 K),[4] research activities to the GICs family have been amplified. Different intercalated elements lead to different interlayer distance. Generally, the smaller the layer-layer separation, the larger the T$_c$.[5] In BaC$_6$, the interlayer distance is so large that the superconductivity is almost suppressed. LiC$_6$ is an exception, in which the small intercalant-graphite distance results in an up-shift of the intercalant band above the Fermi level. As a result, superconductivity in LiC$_6$ is also suppressed.

Although superconductivity in layered structures have been extensively studied, little attention has been paid to two-dimensional (2D) single layers. Notice that 2D monolayer may have different superconducting properties than its corresponding layered bulk materials. For example, different from the non-superconducting bulk LiC$_6$, LiC$_6$ monolayer was predicted to have a T$_c$ about 8 K.[6] Using a rigid band approximation, Savini et al. also predicted p-doped graphane is an electron-phonon mediated superconductor with a T$_c$ above 90 K.[7] A pristine 2D superconducting system without doping is very attractive both fundamentally and practically. If existing, such a simple system will provide us an ideal platform to understand low-dimensional superconductivity, as MgB$_2$ did in three dimensions.[3a]

To find a simple 2D superconductor, we first check the requirements for a material to be a BCS superconductor. In fact, most of the phonon mediated superconductors found after MgB$_2$, including boron-doped diamond,[8] silicon,[9] and silicon carbides,[10] barium-doped silicon clathrates,[11] alkali-doped fullerenes,[12] GICs,[4] and recent predicted p-doped graphane,[7] can be grouped into "covalent-metal".[13] They are metallic yet have strong covalent bonds in their metallic states.

As we know, within the framework of BCS theory, McMillan formula[14] is used to predict T$_c$. It is related to the logarithmically averaged characteristic phonon frequency $\omega_{\ln}^{ph}$, the electron-phonon coupling parameter $\lambda$, and the Coulomb pseudopotential $\mu^*$ which describe the effective electron-electron repulsion.

$$T_c = \frac{\omega_{\ln}^{ph}}{1.20}\exp[\frac{-1.04(1+\lambda)}{\lambda - \mu^*(1+0.62\lambda)}] \qquad (1)$$

Therefore, T$_c$ will be maximized by increasing $\omega_{\ln}^{ph}$ and $\lambda$. Materials composed of light elements will have high frequency phonon modes, which mat increase $\omega_{\ln}^{ph}$. Besides, $\lambda$ is directly related to the electron-phonon coupling potential V and the electronic density of states (DOS) at the Fermi energy ($N(E_F)$), $\lambda = (N(E_F)V$. Larger V and $N(E_F)$ will also lead to a higher T$_c$. With these rules in mind, searching for superconductors is till very challenging, because $\omega_{\ln}^{ph}$, V and $N(E_F)$ are strong interwound. However, a rule of thumb is that phonon mediated superconductivity is more likely to appear in metallic systems with covalent bonds.

In this paper, by using first-principles lattice dynamics and electron-phonon coupling (EPC) calculations, we predicted that B$_2$C single layer is a 2D intrinsic BCS superconductor. The calculated $\omega_{\ln}^{ph}$ and EPC parameter $\lambda$ are 314.8 K and 0.92, respectively. They resulted a T$_c$ of 19.2 K for $\mu^*$ equal to 0.10.

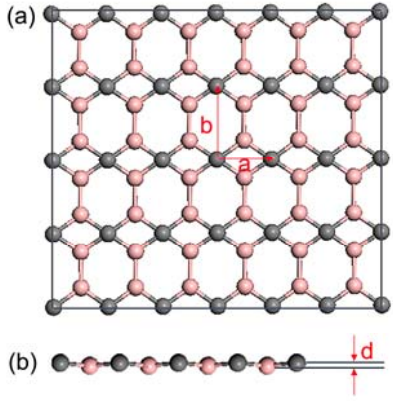

**Fig. 1** (a) Top and (b) side view of optimized B$_2$C monolayer. Gray and pink spheres are carbon and boron atoms, respectively.

Electronic structure calculations and geometrical optimizations are performed using density-functional theory (DFT)[15] within local-density approximation (LDA).[16] Such a protocol has been successfully applied in similar systems, such as boron-doped diamond.[17] A vacuum space around 20 Å along the direction perpendicular to the B$_2$C sheet is used in order to eliminate the interlayer interaction generated by periodic boundary condition. The electron-ion interaction is described by ultrasoft pseudopotentials,[18] and a energy cutoff of 60 Ry is used. EPC calculations are carried out using density-functional perturbation theory with linear response.[19] The $k$-point integration for geometrical optimization, construction of the induced charge density, and calculations of dynamical matrix is performed over a 36×36×1 Monkhorst-Pack grid,[20] and a finer 72×72×1 grid is used in the phonon linewidth calculations, where the convergence in the $k$-point sampling is more difficult. The dynamical matrix and phonon linewidth are computed on a 12×12×1 $q$-point mesh, and a Fourier interpolation is used to obtain complete phonon dispersions. A smearing of 0.02 Ry is used to speed up convergence. The convergence of phonon dispersion with respect to energy cutoff, $k$-points sampling and smearing has been carefully checked, the choice of 60 Ry, 36×36×1 grid and a smearing of 0.02 Ry gives convergence within about 2% on phonon modes at Γ and X and less than 1 mRy on the total energy.

As shown in Fig. 1, B$_2$C monolayer is composed of rectangular B$_2$C building blocks. In the $a$ direction, neighboring rectangular B$_2$C units share two common boron atoms. In the $b$ direction, neighboring B$_2$C units are connected via B-B covalent bonds. By this way, it forms a sheet composed of hexagons and rhombouses. The optimized lattice constants are $a$=2.552 Å, and $b$=3.392 Å. The bond length of B-B and B-C are 1.667 Å and 1.541 Å, respectively. The B$_2$C monolayer is slightly corrugated with the boron layer and carbon layer separated by ~0.032 Å. Although it has not yet

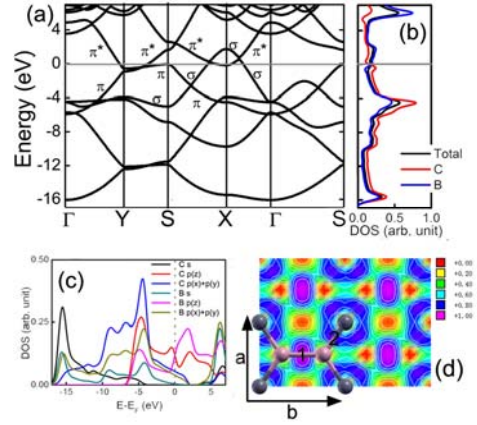

**Fig. 2** (a) Electronic band structure and (b) density of states (DOS), (c) orbital-projected DOS of B$_2$C, (d) electron localization function of B$_2$C projected onto the B-plane. The total DOS in (b) is normalized to per atom for comparison. Γ (0.0, 0.0, 0.0), Y (0.5, 0.0, 0.0), S (0.5, 0.5, 0.0) and X (0.0, 0.5, 0.0) refer to the special points in the first Brouillon zone.

been synthesized, B$_2$C monolayer is predicted to be able to maintain its structural integrity up to 2000 K by Wu et al.[21]

The electronic band structure and DOS are shown in Fig. 2. B$_2$C sheet is metallic with three bands crossing the Fermi level (the bonding π band, the antibonding π* band, and the bonding σ band). The presence of $p_z$-derived π and π* band and $p_{x,y}$-derived σ band at the Fermi level are similar to both MgB$_2$[3b], and GICs,[24b] while the interlayer states in GICs are missing here due to the 2D character. In the atom-projected DOS, we can see both B and C contributions to the states at Fermi level. Orbital decomposed DOS (Fig. 2(c)) is obtained by using Löwdin population analysis. As expected from the band structure, the $p_z$ orbital of C and B dominates at $E_F$, and there is also a smaller $p_y$ ($p_x$) contribution from C and Bσ character. N($E_F$) is 0.56 states/eV per 3-atom unit cell, close to that of MgB$_2$ (0.71 states/eV/cell)[3a] but smaller than CaC$_6$ (1.59 states/eV/unitcell).[24b]

In order to get a detail information about the bonding nature of B$_2$C sheet, we also performed electron localization function (ELF) analusis.[22] ELF is related to the total electron density and its gradient, and it has been demonstrated to be very useful in terms of distinguishing different binding interactions.[23] ELF values ranges from 0 to 1 by definition. For perfectly localized electrons, such as participated in paired covalent bond, the corresponding ELF value should be close to 1. For a free electron gas, the ELF value should be smaller than 0.5. As shown in Fig. 2(d), there is a very strong covalent bond between B atoms, with an ELF value around 1. While the ELF value for B-C bonds is around 0.85, which also indicates a covalent bonding. In the area between two B-B bonds along the $b$ direction, ELF values are approaching 0.5, suggesting delocalized electrons.

Now, we can see that that B$_2$C meets the essential requirements for a BCS-type superconductor, namely, being metallic and covalent bonding. It is thus desirable to directly check its phonon dispersion and electron-phonon coupling property. The phonon band structure and DOS are shown in Fig. 3. First, we are confirmed that B$_2$C is dynamically stable, with a well-converged phonon band structure and no structural instability. Very recently, Xiang et al. have found another
5 low-energy $B_2C$ structure.[24] Optimized with the computational method adopted in this study, their structure is 44.7 meV per atom lower in energy than the current structure. However, we emphasize that, provided it is dynamically stable, even a metastable structure should be able to obtain experimentally
10 in principle.

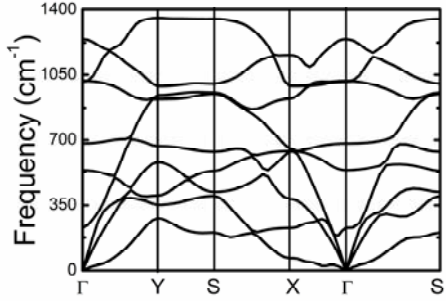

**Fig. 3** Phonon band structure of $B_2C$ sheet.

Based on group theory, the 9 modes at $\Gamma$ points can be decomposed as:
$$\Gamma_{vib} = 3A_1(IR+R) + 3B_1(IR+R) + A_2(R) + 2B_2(IR+R)$$

The calculated frequency of the highest optical mode at $\Gamma$ is
15 1243 cm$^{-1}$, which is corresponding to a B-B stretching mode. Such a high frequency is consistent with the strong covalent bonding suggested by ELF. This mode is active in both Raman and IR spectroscopy.

We then use the Eliashberg function $\alpha^2F(\omega)$ to analyze
20 contribution of each phonon mode to the electron-phonon coupling constant:

$$\alpha^2F(\omega) = \frac{1}{2N_q}\sum_{q\nu}\lambda_{q\nu}\omega_{q\nu}\delta(\omega-\omega_{q\nu}) \quad (2)$$

where $N_q$ is the number of q points used, and $\lambda_{q\nu}$ is the electron-phonon interaction for a phonon $\nu$ with momentum
25 q, which can be written as:

$$\lambda_{q\nu} = \frac{4}{\omega_{q\nu}N(0)N_k}\sum_{k,n,m}\left|g^\nu_{kn,k+qm}\right|^2\delta(\varepsilon_{kn})\delta(\delta_{k+qm}) \quad (3)$$

Here $N(0)$ is the density of states at Fermi energy, and $N_k$ is the number of $k$ points, the matrix element is:

$$g^\nu_{kn,k+qm} = \frac{\langle kn|\delta V/\delta u_{q\nu}|k+qm\rangle}{\sqrt{2\omega_{q\nu}}} \quad (4)$$

30 where $u_{q\nu}$ is the amplitude of the displacement of the phonon and V is the Kohn-Sham potential.

We show in the Fig. 4 the Eliashberg function $\alpha^2F(\omega)$, the integral $\lambda(\omega) = 2\int d\omega' \alpha^2F(\omega')/\omega'$ and the displacement decomposed partial phonon DOS of B and C in in-plane ($xy$,
35 parallel to the $B_2C$ sheet) and out-of-plane ($z$, perpendicular to the $B_2C$ sheet) contributions. The partial phonon DOS for atom $a$ is defined as: $\rho_a(\omega) = \sum_q\sum_{j=1}^{3N}|e_a(q,j)|^2\delta[\omega-\omega(q,j)]$, where $N$ is the total number of atoms, q is the phonon momentum, $j$ labels the phonon branch, $e_a(q,j)$ is the phonon
40 displacement vector for atom $a$, and $\omega(q,j)$ is the phonon frequency. In MgB$_2$ and GICs, due to the weak bonding

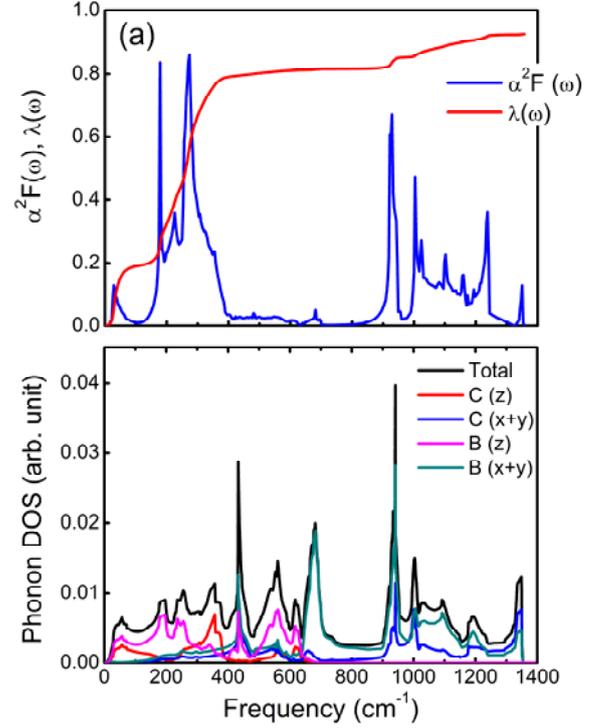

**Fig. 4** (a) Eliashberg spectral function $\alpha^2F(\omega)$ and integrated electron-phonon coupling $\lambda(\omega)$ for $B_2C$ sheet. (b) Total phonon DOS and partial
45 phonon DOS projected on selected vibrations.

between Mg$^{2+}$ (alkali atom) and B plane (graphene), the decomposed phonon DOS is well separated in energy for different elements.[25] In $B_2C$ sheet, the strong bonding between B and C make phonon DOS of B and C largely interwined
50 except for the B out-of-plane mode around 200 cm$^{-1}$ and B in-plane modes around 700 and 1100 cm$^{-1}$.

Different from MgB$_2$ in which the electron-phonon interaction almost comes from the E$_{2g}$ phonon entirely, and that in GICs where symmetry forbids the coupling between $\pi$
55 -states and the softer out-of-plane vibrations, coupling with out-of-plane vibrations is significantly promoted in $B_2C$ sheet. Besides, vibrations in the range from 200 to 400 cm$^{-1}$ contribute the largest part of the total $\lambda$, namely about 0.5 of 0.92. Most of this part of the coupling is from out-of-plane
60 modes, especially, in the region near 200 and 350 cm$^{-1}$. The strong coupling of electrons with out-plane modes is presumedly due to the dominating $\pi$ character, and the coupling of $\pi$ electron with out-of-plane modes in $B_2C$ sheet is symmetry allowed. $\omega^{ph}_{ln}$ of $B_2C$ is 314.8 K, larger than that
65 of CaC$_6$ (286.6 K )[24b], but much smaller than that of MgB$_2$ ($\sim$700 K)[3a,3b]. The typical $\mu^*$ is in the range 0.10-0.15, thus, by using the McMillan formula, the calculated T$c$ is in the range of 19.2 ($\mu^*$=0.10) to 14.3 ($\mu^*$=0.15), indicating $B_2C$ sheet is an intrinsic BCS-type superconductor with a relative high T$_c$.

In summary, we have carried out a first-principles study on the 2D B$_2$C sheet, and the intrinsic phonon mediated superconductivity is predicted with T$_c$ ranges in 14.3-19.2 K. The low frequency out-of-plane phonon modes contribute to most of the coupling. The intrinsic superconductivity in B$_2$C may offer a better platform to investigate the superconductivity in low dimensional systems, and B$_2$C is also a good starting point to obtain high T$_c$ in 2D systems. T$c$ is expected to increase if more $\sigma$ bands are shifted to the Fermi level, since $\sigma$ states are easier to couple with high-frequency phonon modes.

## Acknowledgements

This work is partially supported by NSFC (20933006, 91021004, and 21173202), by CUSF, by the National Key Basic Research Program (2011CB921404), by the Fundamental Research Funds for the Central Universities, by USTCSCC, SCCAS, and Shanghai Supercomputer Center.

## Notes and references

1  J. Nagamatsu, N. Nakagawa, T. Muranaka, and J. Akimitsu, *Nature,* 2001, **410,** 63-64.
2  Y. Kamihara, T. Watanabe, M. Hirano, and H. Hosono, *J. Am. Chem. Soc.,* 2008, **130,** 3296-3297.
3  (a) J. An and W. Pickett, *Phys. Rev. Lett.,* 2001, **86,** 4366. (b) J. Kortus, I. Mazin, K. Belashchenko, V. Antropov, and L. Boyer, *Phys. Rev. Lett.* 2001, **86,** 4656. (c) K. Bohnen, R. Hield and B. Renker, *Phys. Rev. Lett.,* 2001, **86,** 5771. (d) T. Yildirim, O. Gülseren, J. Lynn, C. Brown, T. Udovic, Q. Huang, M. Rogado, K. Regan, M. Hayward, J. Slusky, T. He, M. Hass, P. Khalifah, K. Inumaru and R. Cava, *Phys. Rev. Lett.,* 2001, **87,** 037001. (e) Y. Kong, O. Dolgov, O. Jepsen and O. Andersen, *Phys. Rev. B,* 2001, **64,** 020501.
4  (a) T. Weller, M. Ellerby, S. Saxema, R. Smith and N. Skipper, *Nature Phys.,* 2005, **1,** 39-41. (b) E. Emery, C. Hérold, M. d'Astuto, V. Garcia, Ch. Bellin, J. Marêché, P. Lagrange and G. Loupias, *Phys. Rev. Lett.,* 2005, **95,** 087003.
5  (a) J. Kim, L. Boeri, J. O'Brien, F. Razavi and R. Kremer, *Phys. Rev. Lett.,* 2007, **99,** 027001. (b) A. Gauzzi, S. Takashima, N. Takeshita, C. Terakura, H. Takagi, N. Emery, C. Hérold, P. Lagrange and G. Loupias, *Phys. Rev. Lett.,* 2007, **98,** 067002. (c) R. Smith, A. Kusmartseva, Y. Ko, S. Saxena, A. Akrap, L. Forró, M. Laad, T. Weller, M. Ellerby and N. Skipper, *Phys. Rev. B,* 2006, **74,** 024505.
6  G. Profeta, M. Calandra and F. Mauri, *arXiv e-print*: arXiv:1105.3736v1, 2011.
7  G. Savini, A. Ferrari, F. Giustino, *Phys. Rev. Lett.,* 2010, **105,** 037002.
8  E. A. Ekimov, V. A. Sidorov, E. D. Bauer, N. N. Mel'nik, N. J. Curro, J. D. Thompson and S. M. Stishov, *Nature,* 2004, **428,** 542-545.
9  E. Bustarret, C. Marcenat, P. Achatz, J. Kapčmarčik, F. Lévy, A. Huxley, L. Ortéga, E. Bourgeois, X. Blase, D. Débarre and J. Boulmer, *Nature,* 2006, **444,** 465-468.
10 (a) Z. Ren, J. Kato, T. Muranaka, J. Akimitsu, M. Kriener and Y. Maeno, *J. Phys. Soc. Jpn.,* 2007, **76,** 103710. (b) M. Kriener, Y. Maeno, T. Oguchi, Z. Ren, J. Kato, T. Muranaka and Akimitsu, J.; *Phys. Rev. B,* 2008, **78,** 024517.
11 (a) H. Kawaji, H. O. Horie, S. Yamanaka and M. Ishikawa, *Phys. Rev. Lett.,* 1995, **74,** 1427. (b) K. Tanigaki, T. Shimizu, K. M. Itoh, J. Teraoka, Y. Moritomo and S. Yamanaka, *Nature Mater.,* 2003, **2,** 653-655.
12 (a) A. F. Heberd, M. J. Rosseinsky, R. C. Haddon, D. W. Murphy and S. H. Glarum, *Nature,* 1991, **350,** 600-601. (b) C. M. Varma, J. Zaanen and K. Raghavachari, *Science,* 1991, **254,** 989-992.
13 X. Blase, E. Bustarret, C. Chapelier, T. Klein and C. Marcenat, *Nature Mater.,* 2009, **8,** 375-382.
14 W. L. McMillan, *Phys. Rev.* 1968, **167,** 331.
15 (a) P. Hohenberg and W. Kohn, *Phys. Rev.,* 1964, **136,** B864. (b) W. Kohn and L. J. Sham, *Phys. Rev.,* 1965, **140,** A1133.
16 (a) D. M. Ceperley and B. J. Alder, *Phys. Rev. Lett.,* 1980, **45,** 556. (b) J. P. Perdew and A. Zunger, *Phys. Rev. B,* 1981, **23,** 5048.
17 (a) X. Blase, Ch. Adessi and D. Connétable, *Phys. Rev. Lett.,* 2004, **93,** 237004. (b) H. J. Xiang, Z. Li, J. Yang, J. Hou and Q. Zhu, *Phys. Rev. B,* 2004, **70,** 212504.
18 P. Vanderbilt, *Phys. Rev. B,* 1990, **41,** 7892.
19 P. Giannozzi, S. Baroni, N. Bonini, M. Calandra, R. Car, C. Cavazzoni, D. Ceresoli, G. L. Chiarotti, M. Cococcioni, I. Dabo, A. D. Corso, S. de Gironcoli, S. Fabris, G. Fratesi and R. Gebauer, *J. Phys. Condens. Matter,* 2009, **21,** 395502.
20 H. J. Monkhorst and J. D. Pack, *Phys. Rev. B,* 1995, **13,** 5188.
21 X. Wu, Y. Pei and X. C. Zeng, *Nano. Lett.,* 2009, **9,** 1577-1582.
22 A. Savin, R. Nesper, S. Wenger and T. F. Fäsler, *Angew. Chem. Int. Ed.,* 1997, **36,** 1808-1832.
23 Z. Li, J. Yang, J. G. Hou and Q. Zhu, *Angew. Chem. Int. Ed.,* 2004, **43,** 6479-6482.
24 X. Luo, J. Yang, H. Liu, X. Wu, Y. Wang, Y. Ma, S.-H. Wei, X. Gong, and H. Xiang, *J. Am. Chem. Soc., 2011*, **133,** 16285–16290. Based on our calculations, the B$_2$C structure reported in this reference with the lowest energy has a transition temperature of about 5 K.
25 (a) R. Osborn, E. A. Goremychkin, A. I. Kolesnikov and D. G. Hinks, *Phys. Rev. Lett.* 2001, **87,** 017005. (b) M. Calandra and F. Mauri, *Phys. Rev. Lett.* 2005, **95,** 237002.